\documentclass[final,5p,times,twocolumn]{elsarticle}

\usepackage[T1]{fontenc}
\usepackage[utf8]{inputenc}

\usepackage{graphicx}
\usepackage{amssymb}
\usepackage{amsmath}
\usepackage{amsthm}

\usepackage{textgreek}

\usepackage{multirow}
\usepackage{multicol}
\usepackage{fourier}
\usepackage{booktabs}
\usepackage{threeparttablex}
\usepackage{longtable}
\usepackage{xfrac}
\usepackage{booktabs}
\usepackage{hyperref}

\usepackage{lipsum}

\usepackage{listings}
\usepackage{xcolor}
 
\definecolor{codegreen}{rgb}{0,0.6,0}
\definecolor{codegray}{rgb}{0.5,0.5,0.5}
\definecolor{codepurple}{rgb}{0.58,0,0.82}
\definecolor{backcolour}{rgb}{0.95,0.95,0.92}
 
\lstdefinestyle{mystyle}{
    backgroundcolor=\color{white},   
    commentstyle=\color{codegreen},
    keywordstyle=\color{magenta},
    numberstyle=\tiny\color{codegray},
    stringstyle=\color{codepurple},
    basicstyle=\ttfamily\scriptsize,
    breakatwhitespace=false,         
    breaklines=true,                 
    captionpos=b,                    
    keepspaces=true,                 
    numbersep=5pt,                  
    showspaces=false,                
    showstringspaces=false,
    showtabs=false,                  
    tabsize=2,
    frame=bt,
    language=Python,
}
 
\usepackage{lineno}

\usepackage{subcaption}

\usepackage[colorinlistoftodos,prependcaption]{todonotes}
\usepackage{regexpatch}
\makeatletter
\xpatchcmd{\@todo}{\setkeys{todonotes}{#1}}{\setkeys{todonotes}{inline,#1}}{}{}
\makeatother

\usepackage[b]{esvect}
\newcommand{\vecvec}[1] {\vv{\vv{#1}}}

\biboptions{square}
\biboptions{comma}

\journal{arXiv}

\newcommand{\rr}{\vec{r} - \vec{r}\,'}

\hypersetup{draft}
\begin{document}

\begin{frontmatter}

\title{A general method for computing thermal magnetic noise arising from thin conducting objects}



\author[nbe]{Joonas Iivanainen \corref{cor}}
\ead{joonas.iivanainen@aalto.fi}
\author[nbe]{Antti J. Mäkinen}
\author[nbe]{Rasmus Zetter}
\author[nbe]{Koos C.J. Zevenhoven}
\author[nbe]{Risto J. Ilmoniemi}
\author[nbe]{Lauri Parkkonen}

\address[nbe]{Department of Neuroscience and Biomedical Engineering, Aalto University School of Science, FI-00076 Aalto, Finland}

\cortext[cor]{Corresponding author}


\begin{abstract}
Thermal motion of charge carriers in a conducting object causes magnetic field noise that interferes with sensitive measurements nearby the conductor. In this paper, we describe a method to compute the spectral properties of the thermal magnetic noise from arbitrarily-shaped thin conducting objects. We model divergence-free currents on a conducting surface using a stream function and calculate the magnetically independent noise-current modes in the quasi-static regime. We obtain the power spectral density of the thermal magnetic noise as well as its spatial correlations and frequency dependence. We describe a numerical implementation of the method; we model the conducting surface using a triangle mesh and discretize the stream function. The numerical magnetic noise computation agrees with analytical formulas. We provide the implementation as a part of the free and open-source software package \texttt{bfieldtools}.
\end{abstract}





\end{frontmatter}



\section{Introduction}

\noindent Thermal agitation of charge carriers in a conductor causes a fluctuating voltage and a current referred to as Johnson--Nyquist noise \cite{johnson1928thermal,nyquist1928thermal}. The thermal current fluctuations in the conductor are associated with a magnetic field that interferes with nearby magnetically sensitive equipment and measurements. Thermal magnetic noise can, e.g., limit the performance of sensitive magnetometers operating in conducting shields (e.g., \cite{Varpula1984, nenonen1996thermal, lee2008calculation}), impose constraints on fundamental physics experiments \cite{lamoreaux1999feeble,munger2005magnetic} and cause decoherence in atoms trapped near conducting materials \cite{henkel2005magnetostatic} as well as in high-resolution transmission electron microscopy \cite{uhlemann2015thermal}. It is therefore important to estimate the magnetic noise contribution from nearby conductors when designing sensitive experiments and devices.

Thermal magnetic noise from conductors can generally be calculated either using direct approaches where the field noise is computed from the modeled noise currents and their statistics (e.g., \cite{Varpula1984,nenonen1996thermal,lamoreaux1999feeble,roth_thermal_1998}) or with reciprocal approaches where the noise is obtained by computing the power loss incurred in the material by a known driving magnetic field (e.g., \cite{lee2008calculation,clem1987johnson}). In simple geometries analytical expressions for the magnetic noise can be obtained using either of the two approaches (e.g., \cite{Varpula1984,nenonen1996thermal,lee2008calculation,lamoreaux1999feeble, roth_thermal_1998}). In more complicated geometries, the noise has to be computed numerically. Numerical methods using the reciprocal approach have been used to compute the frequency-dependent magnetic noise (e.g., \cite{uhlemann2015thermal,storm2017ultra,storm2019detection}), while a method using the direct approach has been suggested to compute the low-frequency noise arising from thin conductors  \cite{sandin2010noise}. 

Here, we outline a direct approach to compute the quasi-static frequency-dependent magnetic noise from a conducting object which can be considered as a surface with a small but possibly non-constant thickness. We examine the internal coupling phenomena associated with the surface currents in order to determine the independent modes of the Johnson current \cite{harding1968quantum}. We use a stream-function formalism similar to a previous analytical calculation on an infinite conducting plane \cite{roth_thermal_1998} and to a semi-analytical computation
on a layered grid of square conducting patches \cite{tervo2016noise}. The cross-spectral density of the magnetic noise can be computed based on the current fluctuations of the individual modes described by a set of Langevin equations; the fluctuation amplitudes are given by the equipartition theorem \cite{gillespie1996mathematics}. Examination of the individual modes gives an intuitive picture on the physics that determine the field noise characteristics. 

We present a numerical implementation of the approach which uses a discretization of the stream function on a triangle mesh representing the surface. The implementation is applicable for any conducting surface, including curved ones. We demonstrate computations in example geometries and, where possible, compare the results with analytical formulas for verification. The implementation is freely available as a part of the open-source Python software package \texttt{bfieldtools} (\url{https://bfieldtools.github.io}; \cite{makinen2020magnetic,zetter2020magnetic}). 

\section{Theory}

\noindent We consider the magnetic noise in a frequency range where the macroscopic Johnson thermal noise current is divergence-free ($\nabla \cdot \vec{J} = 0$). In other words, the macroscopic charge density does not fluctuate, but the current fluctuations are due to microscopic thermal motion of charge \cite{roth_thermal_1998}. This allows us to use stream-function expression for the surface current.

\subsection{Stream function and surface current}

\noindent We shortly introduce stream function expression of the surface current and describe how it relates to physical quantities such as power dissipation and inductive energy. Specifically, we assume a thin surface $S$ with conductivity $\sigma(\vec{r})$ and thickness $d(\vec{r})$. A divergence-free surface-current density on $S$ can be expressed with a stream function $\Psi$ (units A/m) as (e.g., \cite{makinen2020magnetic,peeren2003stream, zevenhoven2014conductive})
\begin{equation}
\label{eq:sfrepres}
    \vec{J}(\vec{r}, t) = \nabla_{\|} \Psi(\vec r, t)  \times  \vec n(\vec r),
\end{equation}
where $\vec n(\vec r)$ is the unit surface normal and $\nabla_{\|}$ is the tangential gradient on the surface. We further express the stream function as a linear combination $\Psi(\vec r,t) = \sum_i s_i(t) \psi_i(\vec{r})$, resulting in the current density
\begin{equation}
\label{eq:krepres}
    \vec{J}(\vec{r}, t) = \sum_i s_i(t) \nabla_{\|} \psi_i(\vec{r}) \times  \vec n(\vec r) = \sum_i s_i(t) \vec{k}_i(\vec{r}),
\end{equation}
where $\vec{k}_i(\vec{r}) = \nabla_{\|} \psi_i(\vec{r}) \times  \vec n(\vec r)$ represent spatial patterns of surface-current density (units 1/m) and $s_i(t)$ their time-dependent amplitudes (units A). The magnetic field can be computed from the patterns using the Biot--Savart law
\begin{align}
\vec{B}(\vec{r},t) &= \frac{\mu_0}{4\pi} \int_S \vec{J}(\vec{r}\,',t) \times \frac{\vec{r}- \vec{r}\,'}{|\vec{r}- \vec{r}\,'|^3} dS' \nonumber \\
&=  \sum_i s_i(t) \frac{\mu_0}{4\pi}  \int_S \vec{k}_i(\vec{r}\,') \times \frac{\vec{r}- \vec{r}\,'}{|\vec{r}- \vec{r}\,'|^3} dS' = \sum_i s_i(t) \vec{b}_i(\vec{r}), \label{eq:bfield}
\end{align}
where $\mu_0$ is the vacuum permeability and $\vec{b}_i(\vec{r})$ is the magnetic field from the pattern $\vec{k}_i$ with a unit amplitude.

The instantaneous power dissipation between patterns $\vec{k}_i$ and $\vec{k}_j$ is \cite{makinen2020magnetic,zevenhoven2014conductive}
\begin{align}
\label{eq:instpowdiss}
    P_{ij}(t)  =  s_i(t) s_j(t) \int_S \frac{1}{\sigma(\vec{r}) d(\vec{r})} \vec{k}_i(\vec{r}) \cdot  \vec{k}_j(\vec{r}) dS 
    = s_i(t) s_j(t) R_{ij},
\end{align}
where $R_{ij}$ is the mutual resistance between the patterns. Similarly, the instantaneous inductive energy between the patterns is given by the mutual inductance $M_{ij}$ \cite{makinen2020magnetic,zevenhoven2014conductive}
\begin{align}
\label{eq:magenergy}
    E_{ij}(t)  &= \frac{1}{2} s_i(t) s_j(t) \frac{\mu_0}{4\pi} \int_S \int_S \frac{\vec{k}_i(\vec{r}) \cdot  \vec{k}_j(\vec{r}\,')}{|\rr|} dS dS'  \nonumber\\
    &= \frac{1}{2} s_i(t) s_j(t) M_{ij}.
\end{align}

The amplitudes of the patterns evolve according to a coupled system of equations (\cite{zevenhoven2014conductive}; \ref{app:circuiteq})
\begin{equation}
\label{eq:circuits}
     \mathbf{M} \frac{d}{dt} \mathbf{s}(t) +  \mathbf{R} \mathbf{s}(t) -  \mathbf{e}(t) = 0,
\end{equation}
where $\mathbf{s}$ is a vector containing the pattern amplitudes $\mathbf{s}[i](t) = s_i(t)$, $\mathbf{M}$ and $\mathbf{R}$ are the mutual inductance and resistance matrices with elements $\mathbf{M}[i,j] = M_{ij}$ and $\mathbf{R}[i,j] = R_{ij}$ defined above, and $\mathbf{e}(t)$ gives the electromotive force (emf) that is coupled to the patterns. Equation system ~\eqref{eq:circuits} is analogous to that of coupled RL-circuits, where $\mathbf{s}$ contains the circuit currents. However, we note that circuit quantities such as $\mathbf{M}$ and $\mathbf{R}$ depend on the normalization of the circuit basis functions $\vec{k}_i$, whereas energy quantities such as power dissipation and inductive energy are free of this ambiguity \cite{zevenhoven2014conductive}.

\subsection{Magnetic Johnson--Nyquist noise}

\noindent Next, we investigate how to model the magnetic Johnson--Nyquist noise using the stream-function approach. The thermal current fluctuations are driven by the Johnson emf, which is proportional to a zero-mean Gaussian white noise process \cite{gillespie1996mathematics}. In this context, equations \eqref{eq:circuits} are coupled Langevin equations.

To determine the statistics of the current fluctuations, we apply the equipartition theorem to the system  \cite{gillespie1996mathematics}. According to the theorem, in a thermal bath with temperature $T$ each independent degree of freedom of the system has an average energy of $k_\mathrm{B} T/2$, with $k_\mathrm{B}$ being the Boltzmann constant. The independent degrees of freedom of the system are given by the eigenvectors of $\mathbf{M}$ as they diagonalize the energy \eqref{eq:magenergy}.


We thus look for independent patterns $\vec{\kappa}_i(\vec{r})$ with diagonal $\mathbf{M}$ as linear combinations of $\vec{k}_j(\vec{r})$. We further require that the patterns $\vec{\kappa}_i(\vec{r})$ diagonalize $\mathbf{R}$ so that also the Langevin equations \eqref{eq:circuits} decouple. As the inductance and resistance matrices are symmetric positive-definite for an ordinary conductor \cite{zevenhoven2014conductive}, these independent patterns can be found, for example, by solving a generalized eigenvalue equation \cite{peeren2003stream, golub2012matrix}, i.e., finding an invertible matrix $\mathbf{V}$ such that
\begin{equation} 
\label{eq:geneig}
\mathbf{R} \mathbf{V} = \mathbf{M} \mathbf{V} \boldsymbol{\Lambda} \Leftrightarrow
\left\{
	\begin{array}{ll}
		\mathbf{V}^\mathrm T \mathbf{R} \mathbf{V} = \text{diag}(r_i, \dots, r_N) \\
		\mathbf{V}^\mathrm T \mathbf{M} \mathbf{V} = \text{diag}(l_i, \dots, l_N),
	\end{array}
\right.
\end{equation}
where $\boldsymbol{\Lambda} = \text{diag}(\lambda_1, \dots, \lambda_{N})$ is a diagonal matrix with $\lambda_i=r_i/l_i$. The independent patterns are given by the columns of the invertible but generally non-unitary matrix $\mathbf{V}$ as $\vec{\kappa}_i(\vec{r}) = \sum_j V_{ji} \vec{k}_j(\vec{r})$. 


We can transform Eq.~\eqref{eq:circuits} to the new basis:
\begin{equation}
    \frac{d}{dt} \mathbf{V}^\mathrm T \mathbf{M} \mathbf{V} \mathbf{V}^{-1} \mathbf{s}(t) + \mathbf{V}^\mathrm T \mathbf{R} \mathbf{V} \mathbf{V}^{-1} \mathbf{s}(t) -  \mathbf{V}^\mathrm T \mathbf{e}(t) = 0\,.
\end{equation}
By defining $\tilde{\mathbf{s}}(t) = \mathbf{V}^{-1} \mathbf{s}(t)$ and $\tilde{\mathbf{e}}(t) = \mathbf{V}^\mathrm T \mathbf{e}(t)$, we obtain a set of decoupled Langevin equations
\begin{equation}\label{eq:rlcircuit}
    \frac{d}{dt} \tilde{s}_i(t) + \lambda_i \tilde{s}_i(t) -  \tilde{e}_i(t)/l_i = 0\,.
\end{equation}
Effectively, we now have a number of independent RL-circuits with time constants $\tau_i = l_i/r_i = 1 / \lambda_i$ driven by emfs $\tilde{e}_i(t)$.

The Johnson emf has a white noise (frequency-independent) power spectral density (PSD) $S_{\tilde{e}_i}$ that can be used to solve the PSD of $\tilde{s}_i$ from the decoupled Langevin equation \cite{gillespie1996mathematics}
\begin{equation} \label{eq:jpsd}
    S_{\tilde{s}_i} (\omega) =  \frac{S_{\tilde{e}_i}}{r_i^2} \frac{1}{1 + ( \omega/\lambda_i)^2}\,,
\end{equation}
where $\omega$ is the angular frequency. The average energy \eqref{eq:magenergy} of the $i^\text{th}$ independent degree of freedom is:
\begin{align}
    \langle E_i(t) \rangle =  &= \frac{1}{2} l_i \langle \tilde{s}_i(t)^2 \rangle 
    =\frac{1}{2} l_i \frac{1}{2\pi}\int_{0}^{\infty} S_{\tilde{s}_i} (\omega) d\omega \nonumber \\
    &= \frac{1}{2} l_i \frac{S_{\tilde{e}_i}}{4 r_i^2}\lambda_i=\frac{S_{\tilde{e}_i}}{8 r_i}\,, \label{eq:ave_ene}
\end{align}
where the brackets $\langle \cdot \rangle$ denote the ensemble average. On the other hand, according to the equipartition theorem the average energy is $\langle E_i\rangle = \frac{1}{2} k_\mathrm{B} T$, which can be used together with equation \eqref{eq:ave_ene} to solve the Nyquist formula for the PSD of the Johnson emf:
\begin{equation}
    S_{\tilde{e}_i} (\omega) = 4 k_\mathrm{B} T r_i\,,
\end{equation}
where $r_i$ is associated with the average power dissipation $\langle P_i(t) \rangle = r_i \langle \tilde{j}_i(t)^2 \rangle$.

\begin{figure*}[t]
\centering
\includegraphics[width=1\linewidth]{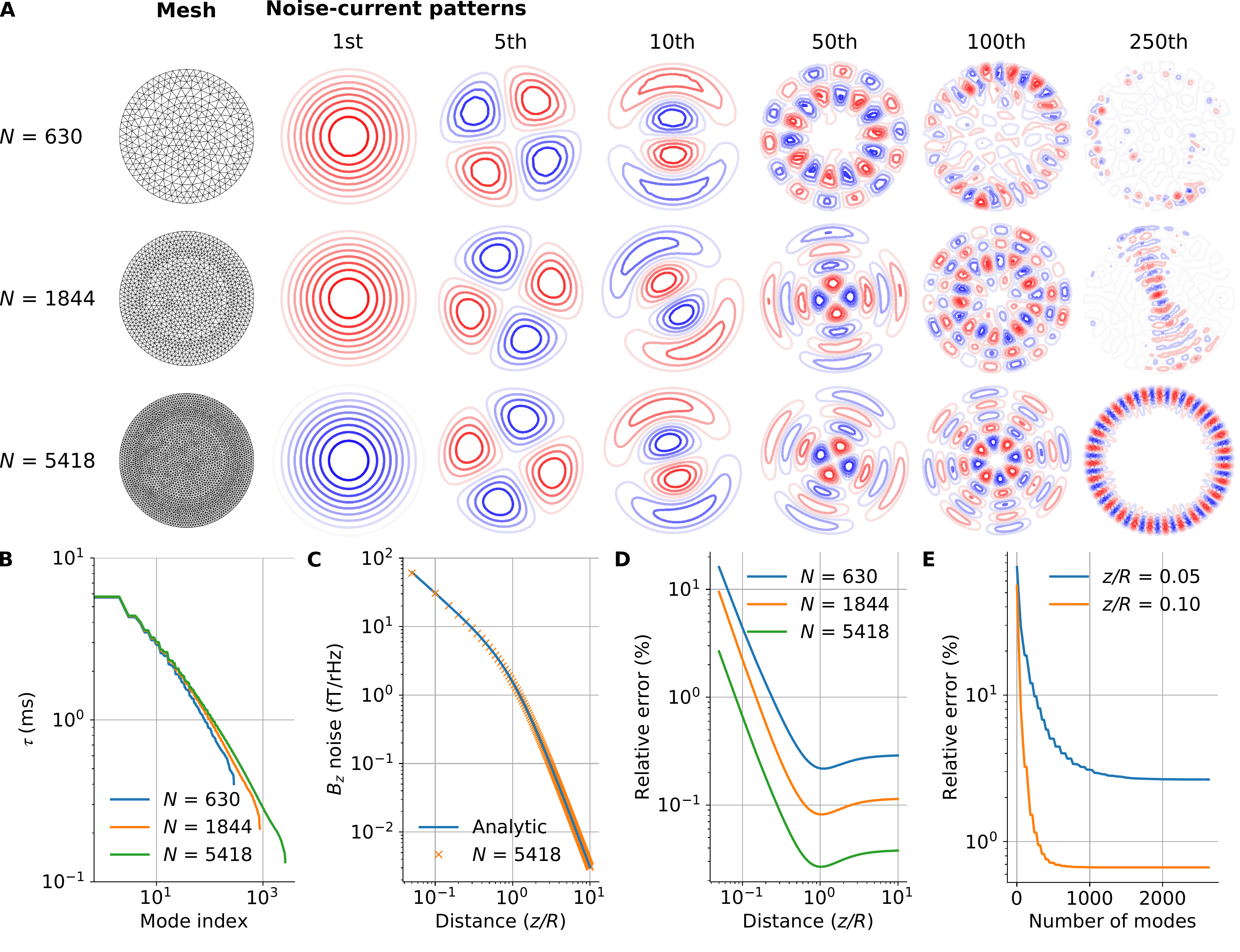}
\caption{Comparison of the numerical solution of magnetic thermal noise of a circular conducting disk with radius $R=1.0$ m (centered on the $xy$-plane) to an analytical formula. {\bf A:} Three meshes with different numbers of triangles ($N$) representing the disk and exemplary contours of the numerically solved noise-current patterns. Blue and red contours depict current flows in opposite directions. {\bf B:} The time constants $\tau$ of the modes computed using the meshes. {\bf C--D:} Comparison between the numerical and analytical solution of the low-frequency magnetic noise ($B_z$) on the $z$-axis. {\bf C:} Comparison between the numerical solution obtained using the densest mesh and the analytical formula. {\bf D:} The relative errors of the numerical solutions to the analytical formula. {\bf E:} Relative error as a function of number of current modes for the densest mesh.}
\label{fig:discvalid}
\end{figure*}

To compute the cross-spectral density (CSD) of the magnetic noise due to the Johnson current, we note that the Fourier transform of the field from the independent patterns is obtained as
\begin{align}
\label{eq:bfieldomega}
    \mathcal{F}\{\vec{B}(\vec{r})\} (\omega) &= \mathcal{F} \big \{ \sum_i s_i(t) \vec{b}_i(\vec{r})\big \} \nonumber \\
    &= \mathcal{F} \big \{ \sum_i \tilde{s}_i(t) \vec{\beta}_i(\vec{r})\big \} =  \sum_i \mathcal{F} \{ \tilde{s}_i \} (\omega) \vec{\beta}_i(\vec{r}),
\end{align}
where $\vec{\beta}_i(\vec{r})$ denotes the magnetic field from $\vec{\kappa}_i$. The CSD between magnetic field components at $\vec{r}$ and $\vec{r}\,'$ along unit vectors $\vec{n}$ and $\vec{n}\,'$ is given by
\begin{align} \label{eq:Bpsd}
  &\Big \langle \vec{n} \cdot \mathcal{F}\{\vec{B}(\vec{r})\}^\ast  \mathcal{F}\{\vec{B}(\vec{r}\,')\}  \cdot \vec{n}\,'  \Big \rangle \nonumber \\
  &= \Big \langle  \vec{n} \cdot  \big ( \sum_i \mathcal{F} \{ \tilde{s}_i \} ^\ast \vec{\beta}_i(\vec{r})  \big ) \big ( \sum_k \mathcal{F} \{ \tilde{s}_k \} \vec{\beta}_k(\vec{r}\,')  \big ) \cdot \vec{n}\,'  \Big \rangle \nonumber \\
    &= \vec{n} \cdot \Big ( \sum_i \sum_k \vec{\beta}_i(\vec{r})   \Big \langle \mathcal{F} \{ \tilde{s}_i \} ^\ast  \mathcal{F} \{ \tilde{s}_k \}  \Big \rangle \vec{\beta}_k(\vec{r}\,') \Big ) \cdot \vec{n}\,' \nonumber \\
    &= \vec{n} \cdot  \vecvec{\mathrm{CSD}}_{\vec{B}}(\vec{r}, \vec{r}\,', \omega) \cdot \vec{n}\,',
\end{align}
where we defined $\vecvec{\mathrm{CSD}}_{\vec{B}} = \sum_i \sum_k \vec{\beta}_i(\vec{r})  \langle \mathcal{F} \{ \tilde{s}_i \} ^\ast  \mathcal{F} \{ \tilde{s}_k \}  \rangle \vec{\beta}_k(\vec{r}\,')$ as the CSD tensor of the magnetic field.

The CSD tensor can be simplified by noting that the amplitudes $\tilde{s}_i$ are independent: their temporal cross-correlation is $\int \tilde{s}_i(t) \tilde{s}_k(t + t') dt = 0$ for $i\neq k$. For $i= k$, the auto-correlation with exponential decay is given as the Fourier transform of the PSD of Eq. \eqref{eq:jpsd}. The CSD of $\tilde{s}_i$ and $\tilde{s}_k$ is thereby $ \langle \mathcal{F} \{ \tilde{s}_i \}^\ast  \mathcal{F} \{ \tilde{s}_k \} \rangle = S_{\tilde{s}_i} (\omega) \delta_{ik}$ and the CSD tensor of the magnetic noise is
\begin{equation}
    \vecvec{\mathrm{CSD}}_{\vec{B}}(\vec{r}, \vec{r}\,', \omega)  = \sum_i \vec{\beta_{i}} (\vec{r})  S_{\tilde{s}_i} (\omega)  \vec{\beta_{i}} (\vec{r}\,').
\end{equation}

We next describe how to compute the CSD between field measurements by an array of sensors. We approximate the  measurement of the $i$th sensor $y_i(t)$ as a weighted sum of the magnetic field over the spatial extent of the sensor
\begin{align}
    y_i(t) =  \int \vec{w}_i(\vec{r}) \cdot \vec{B}(\vec{r},t)  dV
    \approx \sum_{l=1}^{N_i} \vec{w}_i(\vec{r}_{l}) \cdot \vec{B}(\vec{r}_{l},t) , \label{eq:sensor}
\end{align}
where $\vec{r}_{l}$ are the $N_i$ integration points of the sensor $i$ and $\vec{w}_i(\vec{r}_{l})$ are their vector weights. The CSD between measurements $y_i$ and $y_k$ is then
\begin{align}
    \mathrm{CSD}_{y_i,y_k}(\omega) &= \big \langle \mathcal{F} \{ y_i \} ^\ast \mathcal{F} \{ y_k \}  \big \rangle \nonumber \\ &= \sum_{l=1}^{N_i} \sum_{h=1}^{N_k} \vec{w}_i(\vec{r}_{l})^\ast \cdot \vecvec{\mathrm{CSD}}_{\vec{B}}(\vec{r}_{l}, \vec{r}_{h},  \omega ) \cdot \vec{w}_k(\vec{r}_{h}). \label{eq:meascsd}
\end{align}

\section{Implementation}

\noindent In this Section, we briefly outline the numerical implementation of the magnetic noise computation. The implementation is a part of the \texttt{bfieldtools} Python software package \cite{zetter2020magnetic} and uses its stream-function discretization as well as numerical integrals and functions to compute the resistance and inductance matrices. The theoretical and computational aspects of the software are presented in detail in Ref. \cite{makinen2020magnetic}.

In \texttt{bfieldtools}, the conducting surface is represented by a triangle mesh and the stream-function basis $\Psi(\vec r) = \sum_i s_i h_i(\vec{r})$ consists of piecewise linear functions (or "hat functions") $h_i(\vec{r})$. The hat function value is one at the vertex $i$ and zero at other vertices with linear interpolation on the triangle faces. Each of these basis functions represents an elementary current pattern which circulates around the corresponding vertex $i$. The magnetic field is obtained from the stream function $s_i$ with a linear map (Eq. \eqref{eq:bfield}). For example, the $z$-component of the field at $N$ evaluation points is
\begin{equation} \label{eq:linearB}
    \mathbf{b}_{z} = \mathbf{C} \mathbf{s},
\end{equation}
where $\mathbf{C}$ is an $N \times M$ matrix mapping the $M$ vertex-circulating currents ($\mathbf{s}[i] = s_i$) to field component amplitudes at the evaluation points.

The resistance matrix $\mathbf{R}$ (with surface conductivity $\sigma(\vec r) d(\vec r)$ discretized as constant on the triangles) and inductance matrix $\mathbf{M}$ of the elementary current patterns can be computed using the software. In the case of an open mesh, the boundary conditions of the stream function are set as described in Ref. \cite{makinen2020magnetic}. Multiple separate conductors can be handled by computing the inductances between all the patterns and by forming a block resistance matrix comprising the resistance matrices of the individual conductors. 

We decouple the elementary circuits by solving the generalized eigenvalue equation \eqref{eq:geneig} for eigenvalues $\boldsymbol{\Lambda}$ and eigenvectors $\mathbf{V}$ using SciPy \cite{virtanen2020scipy}. We then evaluate the CSD matrix $\boldsymbol{\Sigma}_\mathbf{b}$ of the magnetic field component at $\omega$ using equation \eqref{eq:linearB} as
\begin{equation}
    \boldsymbol{\Sigma}_\mathbf{b} =  \big \langle \mathbf{b}_{z} \mathbf{b}_{z}^\mathrm T \big \rangle = \mathbf{C} \mathbf{V}  \big \langle  \tilde{\mathbf{s}} \tilde{\mathbf{s}}^\mathrm T \big \rangle \mathbf{V}^\mathrm T \mathbf{C}^\mathrm T = \mathbf{C} \mathbf{V} \boldsymbol{\Sigma}_{\tilde{\mathbf{s}}} \mathbf{V}^\mathrm T \mathbf{C}^\mathrm T,
\end{equation}
where $\mathbf{s} =\mathbf{V} \tilde{\mathbf{s}}$ and $\boldsymbol{\Sigma}_{\tilde{\mathbf{s}}}$ is a diagonal matrix with elements $\boldsymbol{\Sigma}_{\tilde{\mathbf{s}}}[i,i] = S_{\tilde{s}_i} (\omega)$ (Eq. \eqref{eq:jpsd}).

We model the measurement $y_i$ in Eq. \eqref{eq:sensor} as $y_i = \mathbf{w}_i^\mathrm T \mathbf{b}_i$, where $\mathbf{w}_i^\mathrm T$ is a row vector comprising the sensor weights and $\mathbf{b}_i = \mathbf{C}_i \mathbf{s}$ is a column vector of the magnetic noise along the directions of the vector weights at the integration points. The elements of the measurement CSD matrix can then be computed as follows
\begin{equation}
\boldsymbol{\Sigma}_\mathbf{y}[i,k]  =  \mathbf{w}_i^\mathrm T \big \langle \mathbf{b}_i \mathbf{b}_k^\mathrm T \big \rangle \mathbf{w}_k = \mathbf{w}_i^\mathrm T \mathbf{C}_i \mathbf{V} \boldsymbol{\Sigma}_{\tilde{\mathbf{s}}}  \mathbf{V}^\mathrm T \mathbf{C}_k^\mathrm T, \mathbf{w}_k.
\end{equation}

In practice, we compute the cross-spectral densities using multidimensional NumPy-arrays \cite{oliphant2015guide} and by summing over the relevant dimensions of the arrays. This way, we can, e.g., compute the cross-spectral density of the magnetic field in 300 observation points at 100 frequencies and store the result in an array with dimensions of $300\times300\times3\times3\times100$.

\section{Validation and numerical results}

\noindent We first computed special case examples that allowed comparing our numerical computation of the magnetic noise to analytical formulas at the low-frequency limit. Specifically, we investigated the following:
\begin{itemize}
    \item $B_z$ noise along the $z$-axis due to a uniform conducting disk centered on the $xy$-plane
    \item $B$ noise at the center of a spherical conducting surface as a function of the sphere radius
    \item $B$ noise at the center of a cylindrical conducting surface along the long axis of the cylinder.
\end{itemize}
The analytical formulas for these three cases can be found in the paper by Lee and Romalis \cite{lee2008calculation}. Besides validation, we present other example computations. Unless stated otherwise, we used $d = 1$ mm and $\sigma = 3.8 \times 10^7$ $\Omega^{-1} \mathrm{m}^{-1}$, corresponding to aluminium at room temperature $T = 293$ K.

\subsection{Validation cases}

\noindent Figure \ref{fig:discvalid} presents the computation of the low-frequency $B_z$ noise along the $z$-axis due to a disk with a radius $R = 1.0$ m centered on $xy$-plane. The disk was modeled with three different meshes with 630, 1 844 and 5 418 triangles. Fig.~\ref{fig:discvalid}A shows examples of the stream-function contours of the numerically computed patterns of the noise current, while Fig.~\ref{fig:discvalid}B shows their time constants. At a relative distance of $z = 0.05 R$, the relative error of the numerical solution of $B_z$ noise to the analytical formula is 2.7\% when the densest mesh is used; with a larger distance, the relative error is smaller. Higher-order modes contribute to the noise when the distance is small 
(Fig. \ref{fig:discvalid}E).

\begin{figure}[t]
\centering
\includegraphics[width=\linewidth]{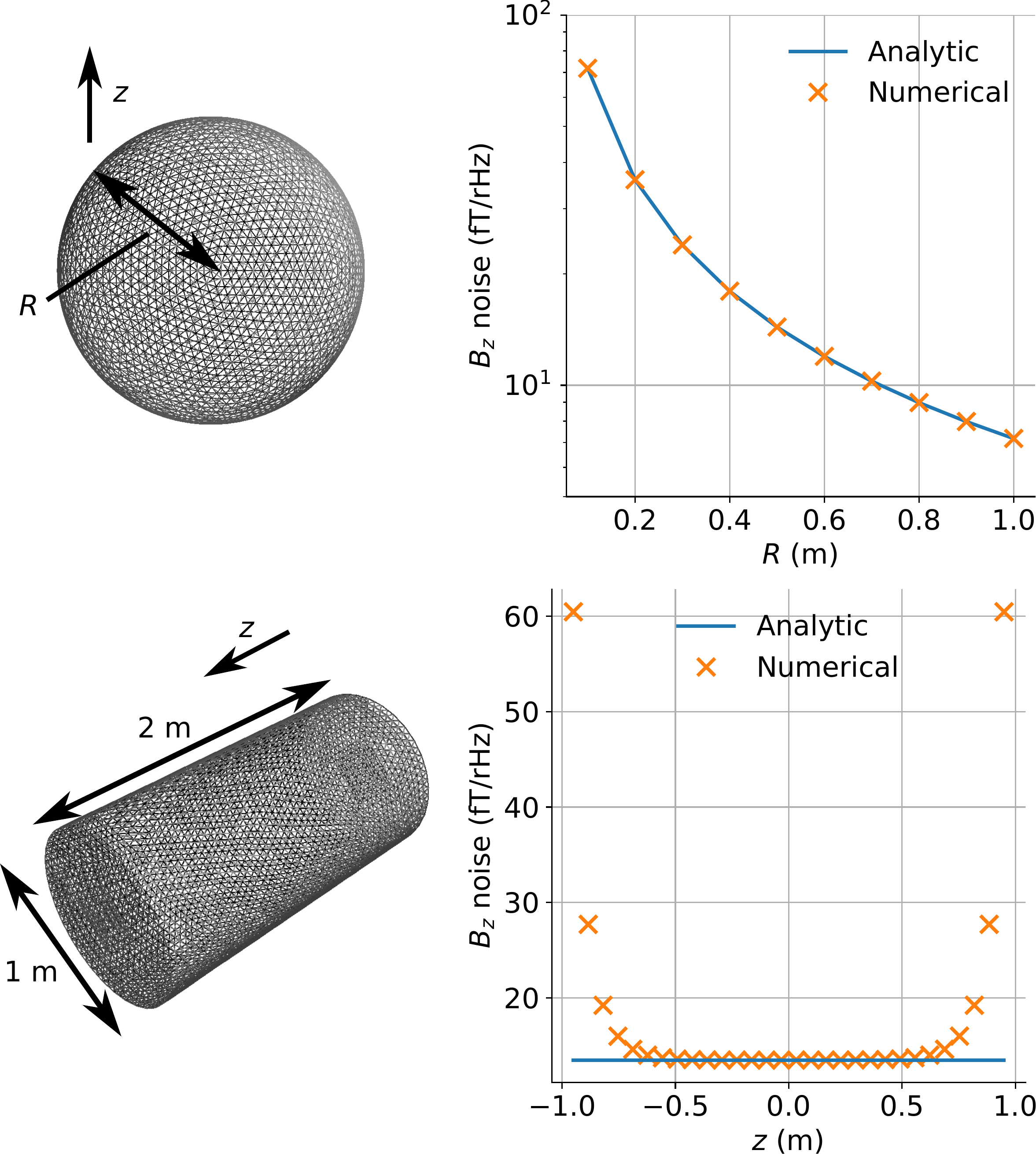}
\caption{Numerical computation of low-frequency magnetic noise inside spherical and cylindrical conducting surfaces (represented as triangle meshes) and comparison to analytical formulas. {\bf Top:} Magnetic noise in the center of the sphere with different sphere radii $R$. {\bf Bottom:} Noise in the field component along the long axis ($z$) of the cylinder. The analytical formula only applies in the center of the cylinder ($z=0$ m).}
\label{fig:valid}
\end{figure}

Figure \ref{fig:valid} shows the numerical results for the low-frequency magnetic noise inside a closed sphere (2 562 vertices; 5 120 triangles) and a cylinder (3 842; 7 680). The computation and analytical formula agree in both cases, with relative errors of 0.06\% and 0.03\% in the case of the sphere and the cylinder, respectively.

\subsection{More examples}

\noindent Next, we examined the magnetic noise and its frequency dependence using a simple conductor. We computed the $B_z$ noise on the $z$-axis as well as the magnetic noise CSD along the $x$-axis due to a circular conducting disk centered on the $xy$-plane ($R = 1$ m; mesh with 5 418 triangles, Fig. \ref{fig:discvalid}A). 

\begin{figure}[!t]
\centering
\includegraphics[width=\linewidth]{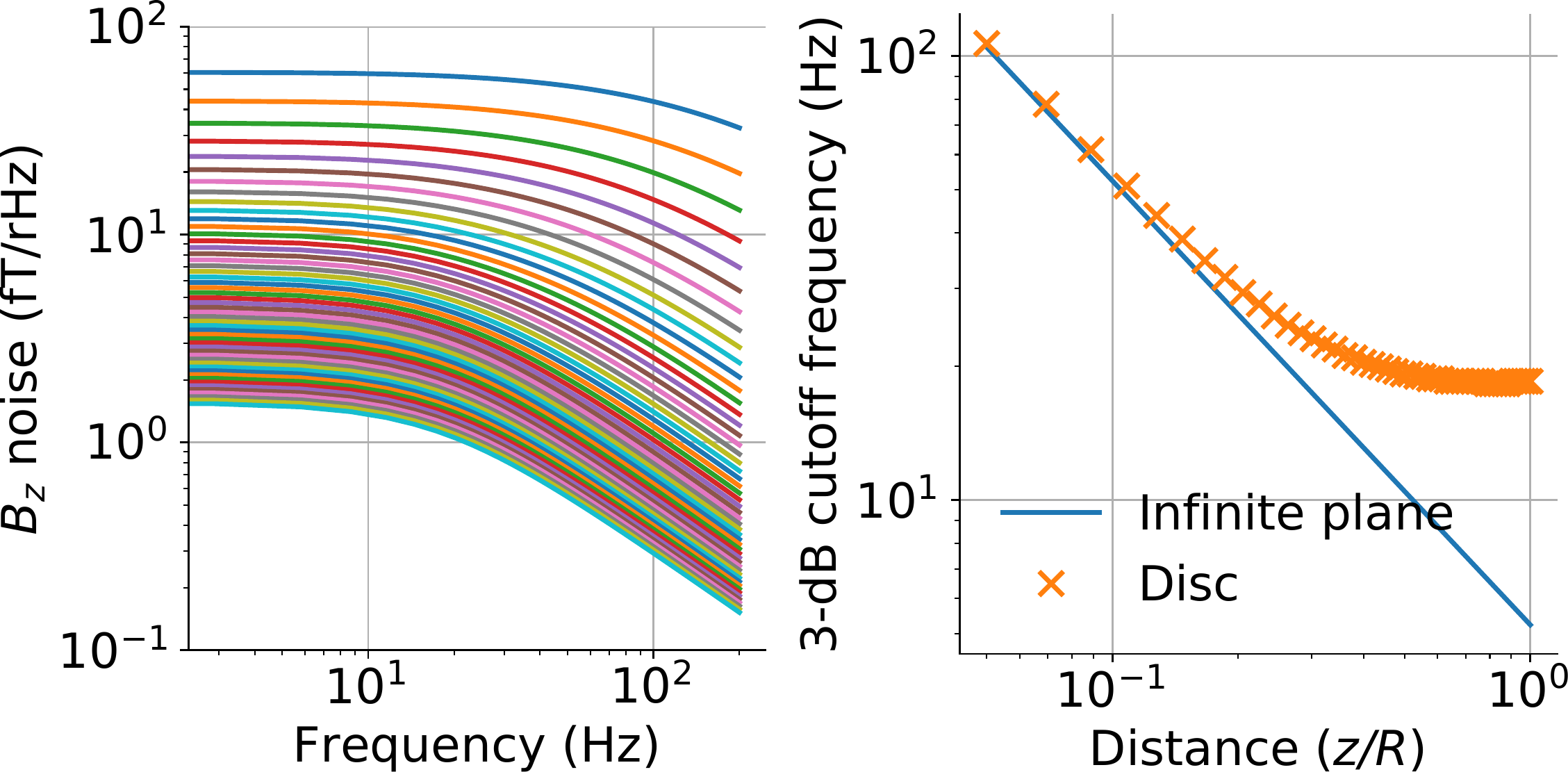}
\caption{The spectral density of the thermal magnetic noise $B_z$ on the $z$-axis due to a circular conducting disk with radius $R=1.0$ m centered on the $xy$-plane. {\bf Left:} Spectral density a function of frequency and distance. The curves with different colors present the noise with different relative distances from the mesh (ranging from $0.05R$ to $R$). {\bf Right:} The frequency at which the noise power has decreased by 3 dB from its zero-frequency value. The solid line gives the cutoff frequencies for an infinite plane calculated using an analytical formula.}
\label{fig:freqvalid}
\end{figure}

\begin{figure}[!t]
\centering
\includegraphics[width=\linewidth]{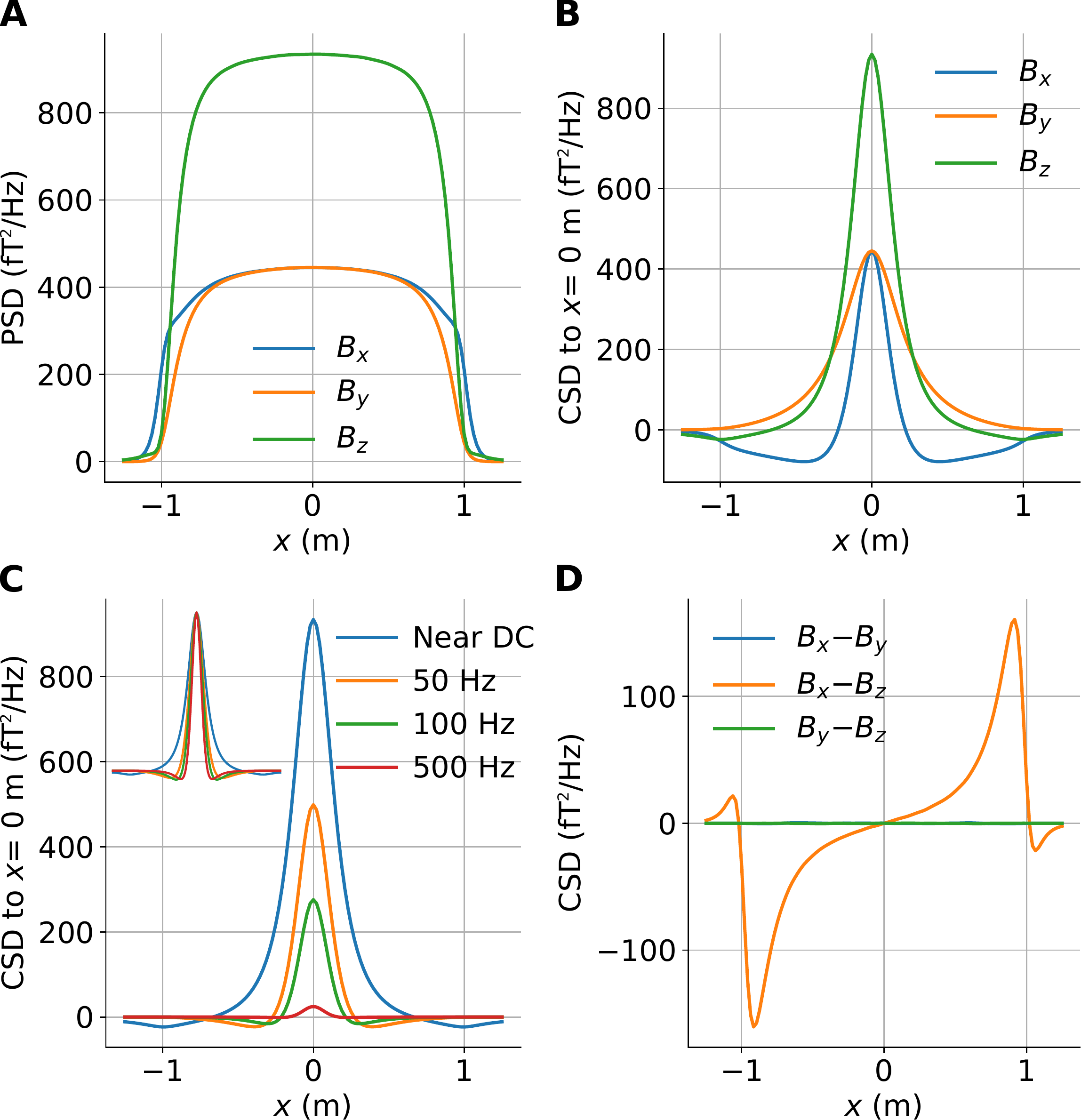}
\caption{Magnetic noise cross-spectral density (CSD) along the $x$-axis ($z = 0.1R$) due to a a circular conducting disk with radius $R=1.0$ m centered on the $xy$-plane. {\bf A:} Low frequency noise power spectral density along the $x$-axis. {\bf B:} Low-frequency noise CSD to $x=0$ m. {\bf C:} Noise cross-spectral density of $B_z$ to $x=0$ m at different frequencies. The inset shows the amplitude-normalized cross-spectral density. {\bf D:} Noise CSD between different components of the magnetic field.}
\label{fig:disc_covar}
\end{figure}

The spectral density of $B_z$ noise due to the disk is shown in Fig. \ref{fig:freqvalid}. The same figure also shows the estimated frequency at which the PSD is reduced by three decibels from the zero-frequency value. The 3-dB cutoff frequency $(4\mu_0 \sigma d z)^{-1}$ for an infinite planar conductor \cite{Varpula1984} is also shown. At small relative distances to the disk ($z < 0.1R$), the numerical 3-dB frequencies scale as those for an infinite plane. At distances comparable to the radius $z \approx R$, the 3-dB frequency is constant suggesting contribution from a single mode with the largest time constant. Fig. \ref{fig:disc_covar} shows examples of cross-spectral density of magnetic noise calculated on the $x$-axis.

\begin{figure*}[!t]
\centering
\includegraphics[width=\linewidth]{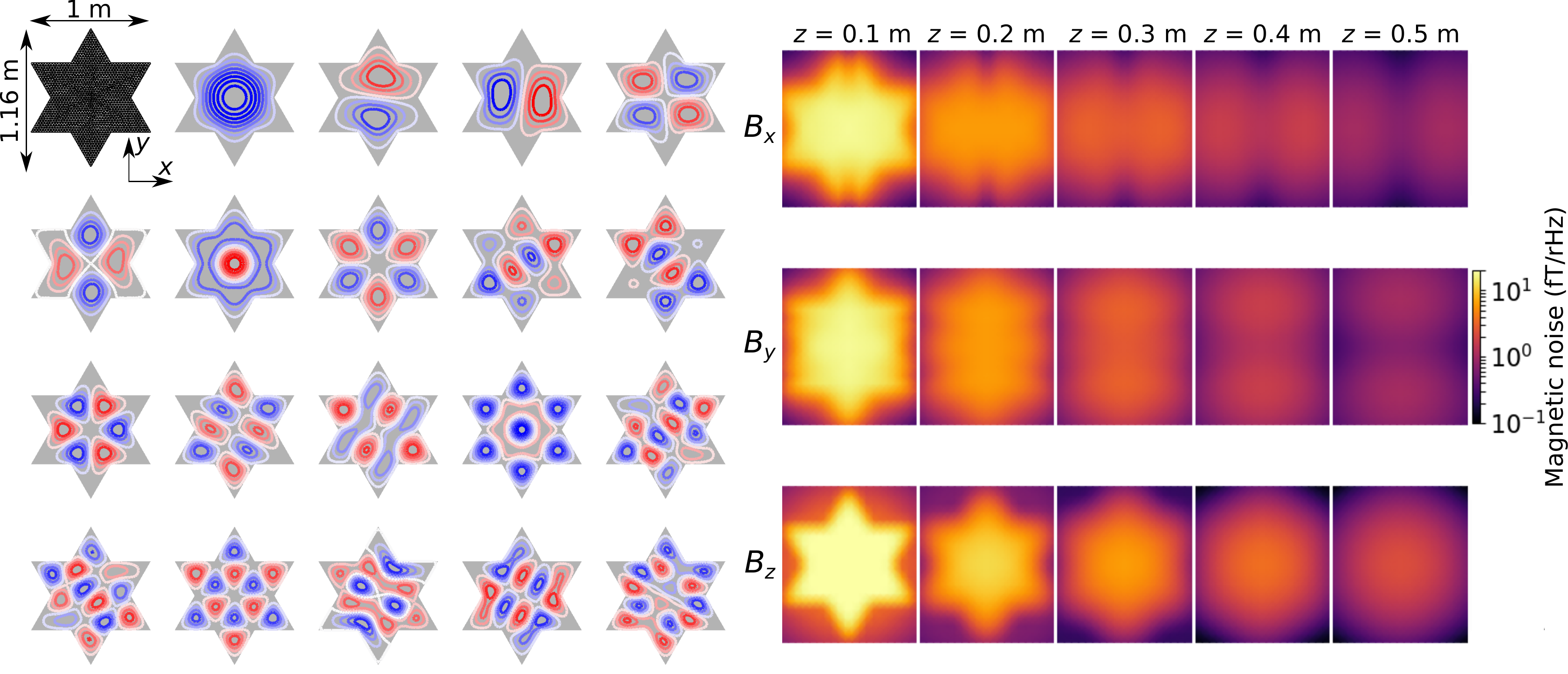}
\caption{Magnetic thermal noise due to a star-shaped planar conductor. {\bf Left:} The triangle mesh representing the conductor and the numerically computed thermal current patterns on the conductor. The time constant of the pattern decreases from left to right, top to bottom. Blue and red contours depict current flows in opposite directions. {\bf Right:} Low-frequency noise spectral density at different vertical distances to the conductor. The plot limits are the same as the size of the conductor shown in A.}
\label{fig:star}
\end{figure*}

We then investigated the magnetic noise due to a planar conductor with a star shape (1 442 vertices, 2 702 triangles). Fig.~\ref{fig:star} illustrates the noise-current patterns on the conductor and the magnetic noise spectral density at different perpendicular distances from the conductor. At small relative distances, the magnetic noise spectral density has a spatial structure that resembles the shape of the conductor. At larger distances, the magnetic noise loses the structural detail, reflecting different fall-off distances of the field noise components that correspond to the noise-current modes with different levels of detail.

\begin{figure*}[!t]
\centering
\includegraphics[width=\linewidth]{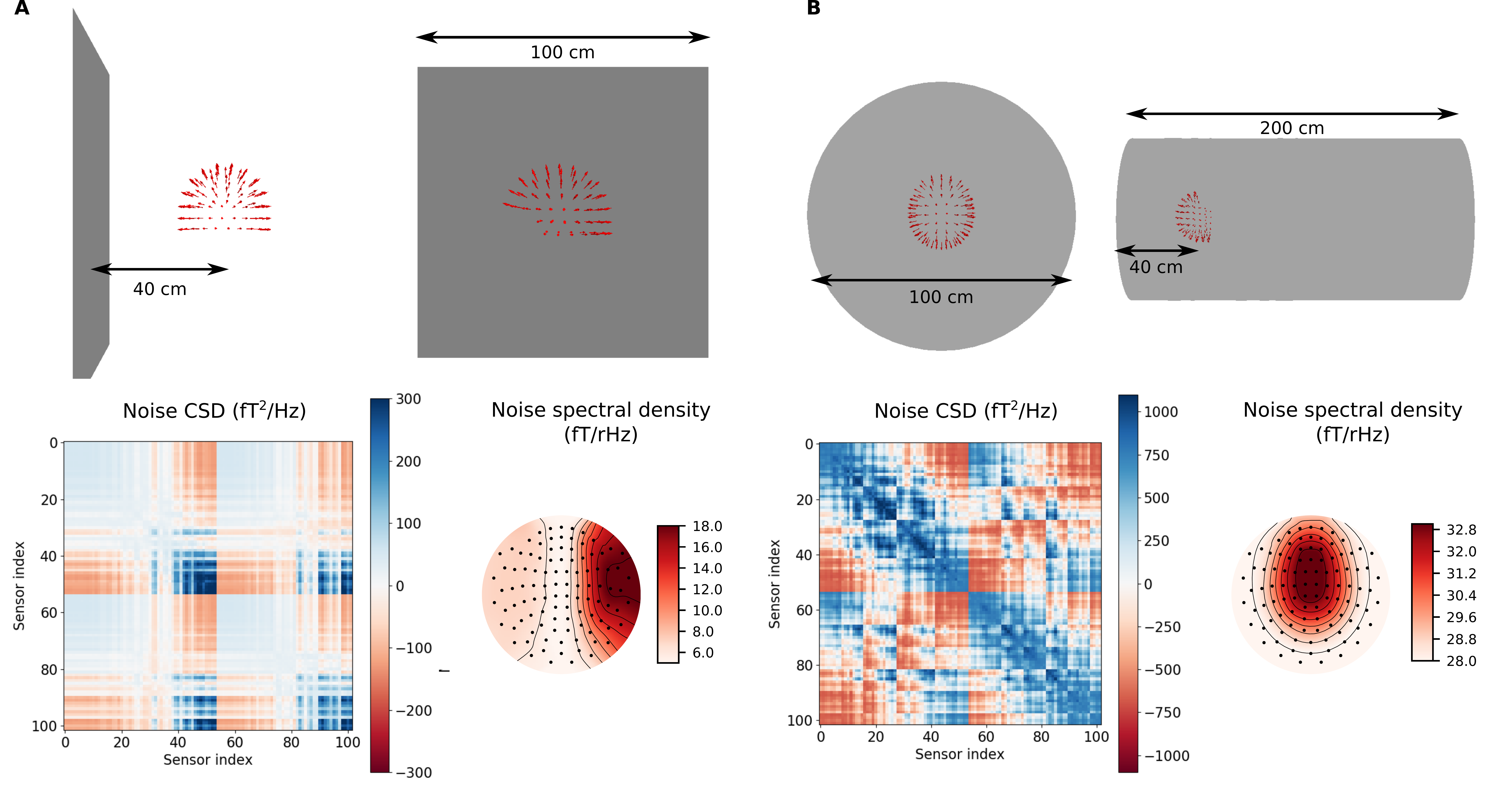}
\caption{Thermal magnetic noise CSD at low frequencies in a SQUID array helmet comprising 102 magnetometers (red arrows) in two geometries: the helmet near an aluminium plate ({\bf A}) and the helmet inside a closed cylindrical aluminium shield ({\bf B}). The noise spectral density is plotted as topographic 2D projection of the sensor-array geometry (generated using the MNE software \cite{gramfort2014mne}). The aluminium is at room temperature and has a thickness of 5 mm in both cases. }
\label{fig:squid}
\end{figure*}

Last, as a practical example, we computed the low-frequency magnetic noise CSD seen by a superconducting quantum interference device (SQUID) array (102 magnetometers; MEGIN Oy, Helsinki, Finland). We investigated two geometries where the magnetometer array was either near an aluminium plate or inside a closed cylindrical aluminium shield (Fig. \ref{fig:squid}). The aluminium had a thickness of 5 mm and was at room temperature in both cases. The magnetometers were modelled as point-like for simplicity.

The computed low-frequency noise spectral CSD in the SQUID array is presented in Fig. \ref{fig:squid}. Compared to the intrinsic noise level of these commercial SQUID sensors ($\sim$3 fT/$\sqrt{\mathrm{Hz}}$), the magnetic noise is significant in both cases.

\section{Conclusion and outlook}

\noindent We presented a method to compute the (cross) spectral density of magnetic thermal noise due to an arbitrarily shaped conductor that can be considered a surface, i.e., thin compared to the distance to the observation points. We have made the implementation openly available (and contributable) as a part of the open-source Python software package \texttt{bfieldtools}. The numerical approach allows visualization of the noise-current patterns, providing an intuitive perspective on the physics. We validated the numerical implementation by comparing the results to analytical formulas, and found agreement within $\sim$1\%. The accuracy increased with the number of triangles in the discretized surface.

We presented examples where we calculated the noise from a single conducting surface, but we note that the computations apply similarly for a system comprising multiple separate conductors that are inductively coupled. Skin effects are not currently included, but could be possibly incorporated by dividing the conductor to multiple inductively-coupled layers each of which has a thickness smaller than the skin depth as in Ref. \cite{sanchez2014multilayer}.

\section*{Acknowledgments}
\noindent This work has received funding from the European Union's Horizon 2020 research and innovation programme under grant agreement No. 820393 (macQsimal). The content is solely the responsibility of the authors and does not necessarily represent the official views of the funding organizations.

\section*{Conflicts of interests}
\noindent The authors declare no conflicts of interest.


\bibliographystyle{ieeetr}
\bibliography{refs.bib}

\appendix

\section{Matrix equation}
\label{app:circuiteq}

\noindent Here, we briefly present the derivation of the matrix equation \eqref{eq:circuits} using the stream-function representation of the surface-current density \eqref{eq:krepres}. We divide the electrical surface current $\vec J(\vec{r},t)$ to two components as
\begin{align}
    \vec J(\vec{r},t) &= \sigma(\vec{r}) d(\vec{r})\vec E(\vec{r},t) = \sigma(\vec{r}) d(\vec{r}) \Big( \vec E_F(\vec{r},t) + \vec E_s(\vec{r},t) \Big ) \nonumber \\
    &=  \sigma(\vec{r}) d(\vec{r}) \Big(-\frac{\partial \vec{A}(\vec{r},t)}{\partial t} + \vec{E}_s(\vec{r},t) \Big ), \label{eq:firsteq}
\end{align}
where $\vec E_F$ is the Faraday-inductive field given as the negative time derivative of the magnetic vector potential $\vec{A}$ and $\vec E_s$ is the source field. The source field represents the active components responsible for the currents in the conductor, while the inductive electric field is due to the magnetic field generated by the currents and represents their inductive coupling. The source field can be, e.g., due to an external time-varying magnetic field ($\vec E_s = - \partial \vec{A}_s / \partial t$) or due to a combination of microscopic thermal current fluctuations $\vec{J}_f$ and their associated macroscopic electric field ($\vec E_s = \vec{J}_f/\sigma d + \vec{E}_f$).

By reordering the terms and expressing the vector potential using the current density, Eq. \eqref{eq:firsteq} reads
\begin{equation}
    \frac{\partial}{\partial t} \frac{\mu_0}{4\pi} \int_S \frac{\vec{J}(\vec{r}\,',t)}{|\rr|} dS'   + \frac{\vec{J}(\vec{r},t)}{\sigma(\vec{r}) d(\vec{r})} - \vec{E}_s(\vec{r},t)= 0.
\end{equation} We consider a frequency range where the charge density does not fluctuate ($\nabla \cdot \vec J = 0$); the current density can be expressed with the stream function:
\begin{equation}
    \frac{\partial}{\partial t}  \sum_k s_k(t) \frac{\mu_0}{4\pi}\int_S \frac{\vec{k}_k(\vec{r}\,')}{|\rr|} dS' +  \sum_k s_k(t) \frac{\vec{k}_k(\vec{r})}{\sigma(\vec{r}) d(\vec{r})} - \vec{E}_s(\vec{r}) = 0.
\end{equation}
By taking a dot product with $\vec{k}_l(\vec{r})$ and integrating over the surface, we have
\begin{align}
    \frac{\partial}{\partial t} & \sum_k s_k(t) \frac{\mu_0}{4\pi} \int_S \int_S \frac{\vec{k}_l(\vec{r}) \cdot \vec{k}_k(\vec{r}\,')}{|\rr|} dS dS' \nonumber \\ 
    &+\sum_k s_k(t) \int_S \frac{\vec{k}_l (\vec{r}) \cdot \vec{k}_k (\vec{r})}{\sigma(\vec{r}) d(\vec{r})} dS - \int_S \vec{k}_l(\vec{r}) \cdot \vec{E}_s(\vec{r},t) dS = 0 ,
\end{align}
from which the resistance and inductance matrix elements can be identified to get the equation system
\begin{equation}
     \sum_k M_{lk} \frac{\partial}{\partial t} s_k(t) + R_{lk} s_k(t) - e_l(t) = 0,
\end{equation}
where $e_l(t) = \int_S \vec{k}_l(\vec{r}) \cdot \vec{E}_s(\vec{r},t) dS$ is the source emf coupled to pattern $l$.

\end{document}